\documentclass[aps,a4paper,showpacs,prc,floatfix,twocolumn,amsmath,amssymb,nofootinbib]{revtex4}
\usepackage{graphicx}
\usepackage{epsfig}
\usepackage{ulem}
\usepackage{morefloats}
\usepackage{rotating}
\usepackage{xcolor,ulem}
\usepackage{graphicx}

\newcommand{\comment}[1]{\textcolor{black}{#1}}

\begin{document}

\title{Virial Expansion of the Electrical Conductivity of Hydrogen Plasmas}

 \author{G. R\"{o}pke}
 \email{gerd.roepke@uni-rostock.de}
 \affiliation{Institut f\"{u}r Physik, Universit\"{a}t Rostock, D-18051 Rostock, Germany}
 \author{M. Sch\"{o}rner}
 \email{maximilian.schoerner@uni-rostock.de}
 \affiliation{Institut f\"{u}r Physik, Universit\"{a}t Rostock, D-18051 Rostock, Germany}
 \author{M. Bethkenhagen}
 \email{mandy.bethkenhagen@ens-lyon.fr}
\affiliation{\'{E}cole Normale Sup\'{e}rieure de Lyon, Laboratoire de G\'{e}ologie de Lyon, LGLTPE UMR 5276, Centre Blaise Pascal, 46 all\'{e}e d'Italie, Lyon 69364, France}
 \author{R. Redmer}
 \email{ronald.redmer@uni-rostock.de}
 \affiliation{Institut f\"{u}r Physik, Universit\"{a}t Rostock, D-18051 Rostock, Germany}

\date{\today}

\begin{abstract}
The low-density limit of the electrical conductivity $\sigma (n,T)$ of hydrogen as the simplest ionic plasma is presented as function of temperature $T$ and mass density $n$ in form of a virial expansion of the resistivity. 
Quantum statistical methods yield exact values for the lowest virial coefficients which serve as benchmark for analytical approaches to the electrical conductivity as well as for numerical results obtained from \comment{density functional} theory based molecular dynamics simulations (DFT-MD) or path-integral Monte Carlo (PIMC) simulations. 
While these simulations are well suited to calculate $\sigma (n,T)$ in a wide range of density and temperature, in particular for the warm dense matter region, they become computationally expensive in the low-density limit, and virial expansions can be utilized to balance this drawback. 
We present new results of DFT-MD simulations in that regime and discuss the account of electron-electron collisions by comparing with the virial expansion. 
\end{abstract}

\maketitle

\section{Introduction.}
Besides the equation of state and the optical properties, the direct-current electrical conductivity $\sigma$ is a fundamental characteristic of plasmas which is relevant in various fields. Examples for technical applications reach from the quenching gas in high-power circuit breakers~\cite{Franck2006} which acts as an efficient dielectric medium up to fusion plasmas produced via magnetic~\cite{Kikuchi} or inertial  confinement~\cite{Lindl}. The electrical conductivity is indispensible for the verification of the insulator-to-metal transition in warm dense hydrogen~\cite{Weir1996}. In geophysics, the electrical conductivity determines the properties of the outer liquid core and of the ionosphere, i.e., the entire magnetic field of Earth from the dynamo region~\cite{RMP2000} up to the magnetosphere~\cite{Kallenrode2004}. Similarly, the electrical conductivity in the convection zone of giant planets~\cite{French2012}, brown dwarfs~\cite{Becker2018}, and stars~\cite{Dynamo} determines the action of the dynamo that produces their magnetic field. The investigation of the electrical conductivity of charged particle systems is, therefore, an emerging field of quantum statistics. In this work we provide exact benchmarks for this fundamental transport property.

Theoretical approaches to calculate the electrical conductivity of plasmas have been performed first within kinetic theory~\cite{landau10}. 
In a seminal paper~\cite{Spitzer53}, Spitzer and H{\"a}rm determined $\sigma$ of the fully ionized  plasma solving a Fokker-Planck equation.
However, to calculate $\sigma (n,T)$ in a wide region of temperature $T$ and mass density $n$, a quantum statistical many-particle theory is needed which describes screening, correlations, and degeneracy effects in a systematic approach. In a very general way, according to the fluctuation-dissipation theorem, the conductivity is expressed in terms of equilibrium correlation functions. Kubo's fundamental approach~\cite{Kubo66} relates the electrical conductivity to the current-current correlation function. For the relation between generalized linear response theory~\cite{Roep88,RR89,Redmer97} and kinetic theory, see~\cite{Reinholz12} and references therein. The evaluation of the corresponding equilibrium correlation functions 
can be performed by using different methods:\\
(i) Analytical expressions are derived, e.g., by using thermodynamic Green's functions. Perturbation theory allows partial summations using diagram techniques which leads to sound results in a wide range of $T$ and $n$. However, as characteristic for perturbative approaches, exact results can be found only in some limiting cases.\\
(ii) This drawback is removed by numerical {\it ab initio} simulations of the correlation functions applicable for arbitrary interaction strength  and degeneracy. Using \comment{density functional} theory (DFT) for the electron system and molecular dynamics (MD) for the ion system, see~\cite{Kubo66,Greenwood,Desjarlais02,Mazevet05,Holst11}, single electron states are calculated solving the Kohn-Sham equations for a given configuration of ions. The total energy is given by the kinetic energy of a non-interacting reference system, the classical electron-electron interaction, and an exchange-correlation energy which contains all unknown contributions in certain approximation. 
One of the shortcomings of this approach is that the many-particle interaction is replaced by this mean-field potential.\\
(iii) In principle, an exact evaluation of the equilibrium correlation functions is possible by using path-integral Monte Carlo (PIMC) simulations, see~\cite{Dornheim2018,PIMC1,PIMC2} and references therein. 
The shortcomings of this approach are the rather small number of particles (few tens), the sign problem for fermions, and the computational challenges to calculate path integrals accurately.

These approaches and other closely related methods have been used to calculate $\sigma(n,T)$ in a wide parameter range, and numerous results have been published, \comment{for a recent review see Ref. \cite{Starrett20}. Also recently, a comparative study~\cite{Grabowski} considering different approaches has been published which revealed large differences of calculated conductivities.} 

\comment{In the present study, we} demonstrate that the virial expansion of the inverse conductivity serves as an exact benchmark for theoretical approaches so that the accuracy and consistency of results for the conductivity~\cite{Grabowski} can be checked. \comment{In particular, we apply this framework to analytical approaches, DFT-MD results, and experimental data for hydrogen, which was chosen for simplicity. In the course of this discussion, we present new DFT-MD data to extend the previously available conductivity data~\cite{Lambert11,Desjarlais} in the density-temperature region of interest.} 
\comment{The virial expansion of $\rho=1/\sigma$ suggested in this work is a prerequisite to work out interpolation formulas for the conductivity. It} can be used in a wide range of $T$ and $n$\comment{; analogous to} 
the Gell-Mann--Brueckner result for the virial expansion of the plasma equation of state, see~\cite{KKER}.
Finally, the benchmark \comment{capability} of the virial expansion \comment{as discussed in this work} may serve as a criterion to check the accuracy of numerical approaches 
like DFT-MD simulations to evaluate the conductivity.

\section{Virial expansion of the inverse conductivity.}  
Charge-neutral hydrogen plasma (ion charge $Z=1$) in thermodynamic equilibrium is characterized by temperature $T$ and the mass density $n$, or the total particle number densities of electrons $\hat n_e$ which equals that of the ions $\hat n_{\rm ion}$. Instead, dimensionless parameters can be introduced: the plasma parameter 
\begin{equation}
\Gamma = \frac{e^2}{4\pi\epsilon_0 k_BT} \left(\frac{4\pi}{3} \hat n_e\right)^{1/3}
\end{equation}
which characterizes the ratio of potential to kinetic energy in the non-degenerate case, and the electron degeneracy parameter 
\begin{equation}
\Theta = \frac{2m_e k_BT}{\hbar^2} (3\pi^2 \hat n_e)^{-2/3}.
\end{equation}
The dc conductivity $\sigma(n,T)$ is usually related to a dimensionless function $\sigma^*(n,T)$ according to 
\begin{eqnarray}
 \sigma(n,T) &=& \frac{(k_BT)^{3/2} (4\pi\epsilon_0)^2}{m_e^{1/2} e^2}\;\sigma^*(n,T) \nonumber\\
 &=&\frac{32405.4}{\Omega {\rm m}}\left(\frac{k_BT}{\rm eV}\right)^{3/2} \sigma^*(n,T) \,.
 \label{eq:1}
\end{eqnarray}
In this work, we consider both $\sigma$ and $\sigma^*$ as function of density $n$ at {\it fixed} temperature $T$. In the low-density limit, the following virial expansion for the inverse conductivity $\rho^*(n,T)=1/\sigma^*(n,T)$ was obtained from kinetic theory and generalized linear response theory~\cite{Roep88,RR89,Redmer97}: 
\begin{equation}
 \rho^*(n,T) = \rho_1(T) \ln\frac{1}{n} + \rho_2(T) + \rho_3(T)\,n^{1/2}\,\ln\frac{1}{n} + \dots
 \label{eq:5}
\end{equation}

In contrast to a simple expansion in powers of $n$, the occurrence of terms with $\ln n$ and $n^{1/2}\,\ln n$ is due to the long-range character of the Coulomb interaction. To describe the collisions between the charged particles, an integral over the Coulomb interaction occurs which gives the so-called Coulomb logarithm, where screening of the Coulomb interaction is taken into account. 
Typically, such a Coulomb logarithm arises in the correlation functions within the generalized linear response theory
~\cite{Roep88,RR89,Redmer97}. 

\comment{By convention, virial expansions consider the dependence of physical quantities on the density $n$, for instance a power series expansion. However the density $n$ has a dimension, and for $\rho^*$ to be not depending on units, 
the virial coefficients $\rho_i$ have also in general a dimension.
In particular, the term  $\rho_1\ln(1/n)$ needs a compensating term $\rho_1\ln(A)$, where $A$ has the dimension of density, as a contribution to $\rho_2$ so that $\rho^*$ remains dimensionless. 
Usually relations like (\ref{eq:5}) are given after fixing the units in which the physical quantities are measured, but it is also convenient to introduce dimensionless variables. For motivation, we consider the Born approximation for the Coulomb logarithm.}

Within static (Debye) screening of the Coulomb interaction to avoid the divergence owing to distant collisions, 
the Born approximation of the Coulomb logarithm leads to the result, see \cite{Roep88,RR89,Redmer97},
\begin{eqnarray}
 && \int_0^\infty \frac{x}{(x+\kappa_{\rm Debye}^2)^2}e^{-\hbar^2x/(8 m_ek_BT)}dx \nonumber \\ 
 && = \ln \left(\frac{\Theta}{\Gamma}\right) - 0.962203 + {\cal O} 
    \left[\frac{\Gamma}{\Theta} \ln \left(\frac{\Theta}{\Gamma}\right)\right] \,. 
 \label{BH}
\end{eqnarray}
The Debye screening parameter in the low-density (nondegenerate) limit reads 
\begin{equation}
\kappa_{\rm Debye}^2 = 2 \hat n_e \frac{e^2}{\epsilon_0 k_BT}
\end{equation} 
so that the integral depends only on the parameter 
\begin{equation}
\frac{\hbar^2 \kappa_{\rm Debye}^2}{8 m_e k_BT} = \left(\frac{2}{3\pi^2}\right)^{1/3} \frac{\Gamma}{\Theta}.
 \end{equation}
We focus on the first and second term on the right hand side of Eq.~(\ref{BH}) which is sufficient in order to derive the first [$\rho_1(T)$] 
and second virial coefficient [$\rho_2(T)$] of the virial expansion (\ref{eq:5}). Further contributions are of higher order in density; for $\Gamma/\Theta \le 0.01$ they contribute to the integral in Eq.~(\ref{BH}) by less than 1~\%.

In the virial expansion (\ref{eq:5}), the logarithm can be transformed by introducing the dimensionless parameter 
\begin{equation}
\frac{\Theta}{\Gamma}=\frac{2m}{\hbar^2} \frac{(k_BT)^2}{\hat n_e} \frac{4 \pi \epsilon_0}{e^2} (36 \pi^5)^{-1/3}\,,
\label{GamT}
\end{equation}
see Eq.~(\ref{BH}), and we find a modified expression [note that $T \propto 1/(\Gamma^2\Theta)$]:
\begin{eqnarray}
 \rho^*(n,T)= \tilde\rho_1 (T) \ln \left(\frac{\Theta}{\Gamma}\right) + \tilde\rho_2 (T) + \dots \,.
 \label{eq:vir}
\end{eqnarray}
\comment{To find the relation between $\tilde\rho_i$ and $\rho_i$ we replace in Eq.~(\ref{eq:vir}) the variables $\Theta,\Gamma$ by $n,T$ according to
Eq.~(\ref{GamT}) so that
\begin{eqnarray}
 \rho^* &=& \tilde\rho_1 (T) \ln \left(\frac{1}{n}\right)\nonumber \\ 
 && + \tilde\rho_1 (T) \ln \left(\frac{2m}{\hbar^2} (k_BT)^2 
 \frac{4 \pi \epsilon_0}{e^2} (36 \pi^5)^{-1/3}\right) \nonumber\\ 
 && + \tilde\rho_2 (T) + \dots 
 \end{eqnarray}
Comparing with Eq.~(\ref{eq:5}) we find $\tilde\rho_1=\rho_1$ and 
\begin{equation} 
 \tilde\rho_2=\rho_2 + \rho_1 \ln[2 \pi (6 \pi)^{2/3} a_B^3 /T_\textrm{Ryd}^2] \,,
\end{equation}
where $a_B$ is the Bohr radius and $T_\textrm{Ryd}=k_BT/{13.6~\textrm{eV}}$ is the temperature measured in Rydberg units. 
}


A highlight of plasma transport theory is that the exact value of the first virial coefficient is known for Coulomb systems from the seminal paper of Spitzer and H{\"a}rm~\cite{Spitzer53},
\begin{equation}
\label{eq:Spitzer}
 \rho_1 = \tilde\rho_1 =\rho^{\rm Spitzer}_1 = 0.846~,
\end{equation}
which does not depend on $T$. 
Note that Eq.~(\ref{eq:Spitzer}) takes into account the contribution of the electron-electron ($e-e$) interaction. In contrast, for the Lorentz plasma model where the $e-e$ collisions are neglected so that only the electron-ion interaction is considered, the first virial coefficient amounts to
\begin{equation}
\rho_1^{\rm Lorentz} =\frac{1}{16} (2\pi^3)^{1/2}=0.492126\,.
 \label{Lorentz}
\end{equation}
Although $e-e$ collisions do not contribute to a change of the total momentum of the electrons because of momentum conservation, the distribution in momentum space is changed ("reshaping") so that higher moments of the electron momentum distribution are not conserved. The indirect influence of $e-e$ collisions on the dc conductivity is clearly seen in generalized linear response theory where these higher moments are considered; see~\cite{Redmer97}.

For the second virial coefficient $\rho_2(T)$ or $\tilde\rho_2(T)$, no exact value is known. 
It depends on the treatment of many-particle effects, in particular screening of the Coulomb potential. 
Within a quantum statistical approach, the static (Debye) screening by electrons and ions, see Eq.~(\ref{BH}), 
should be replaced by a dynamical one. 
For hydrogen plasma as considered here, the Born approximation for the collision integral is justified at high temperatures 
$T_\textrm{Ryd}\gg 1$. Considering screening in the random-phase approximation 
leads to the quantum Lenard-Balescu (QLB) expression. 
Thus, at very high temperatures where the dynamically screened Born approximation becomes valid, 
we obtain the QLB result, see~\cite{Desjarlais,Karachtanov16,foot1},
\begin{equation}
 \lim_{T \to \infty}\tilde\rho_2(T) = \tilde \rho_2^{\rm QLB} = 0.4917~.
 \label{virLB}
\end{equation}

With decreasing $T$, strong binary collisions (represented by ladder diagrams) become important which have to be treated beyond the Born approximation when calculating the second virial coefficient $\tilde \rho_2(T)$. According to Spitzer and H{\"a}rm~\cite{Spitzer53}, 
the classical treatment of strong collisions with a statically screened potential gives for $\rho^*=1/\sigma^*$ the result 
\begin{equation}
\rho_{\rm Sp}^*=0.846 \ln \left[\frac{3}{2} \Gamma^{-3} \right]\,.
\end{equation}
Interpolation formulas have been proposed connecting the high-temperature limit $\tilde\rho_2^{\rm QLB}$ with the low-temperature 
Spitzer limit. 
Instead, performing the sum of ladder diagrams with the dynamically screened Coulomb potential,
Gould and DeWitt~\cite{GDW} and Williams and DeWitt~\cite{WDW} proposed approximations where the lowest order of a ladder sum 
with respect to a statically screened potential, the Born approximation, is replaced by the Lenard-Balescu result which accounts 
for dynamic screening. An improved version was proposed in Refs.~\cite{RR89,RRMK89} by introducing an effective screening 
parameter $\kappa^{\rm eff}$ such that the Born approximation coincides with the Lenard-Balescu result, 
see also~\cite{Redmer97,Roep88,EssRoep98,RR89}. 
Based on a T-matrix calculation in quasiclassical (Wentzel-Kramers-Brillouin, WKB) approximation~\cite{Esser03,RRT89}, 
the expression (temperature is given in eV: $T_{\rm eV}=k_BT/$eV)
\begin{equation}
 \tilde\rho_2(T_{\rm eV}) \approx 0.4917 + 0.846 \ln\left[ 
 \frac{1 + 8.492/T_{\rm eV}}{1 + 25.83/T_{\rm eV} + 167.2/T_{\rm eV}^2} \right]
 \label{WKB}
\end{equation}
can be considered as simple interpolation which connects the QLB result with the Spitzer limit in WKB approximation. 
However, the exact analytical form of the temperature dependence of the second virial coefficient $\tilde\rho_2(T)$ remains an open problem. 

Thus, the available exact results for the virial expansion (\ref{eq:vir}) 
of the resistivity of 
hydrogen plasma are:\\ 
(i) the value of the first virial coefficient is $\tilde\rho_1 = 0.846$; \\
(ii) the second virial coefficient has the high-temperature limit $\lim_{T \to \infty}\tilde\rho_2(T) = 0.4917$; \\
(iii) the second virial coefficient is temperature dependent, a promising functional form is given by Eq.~(\ref{WKB}).

\begin{figure}[htb]
\begin{center}
\centerline{\includegraphics[width=\columnwidth,angle=0]{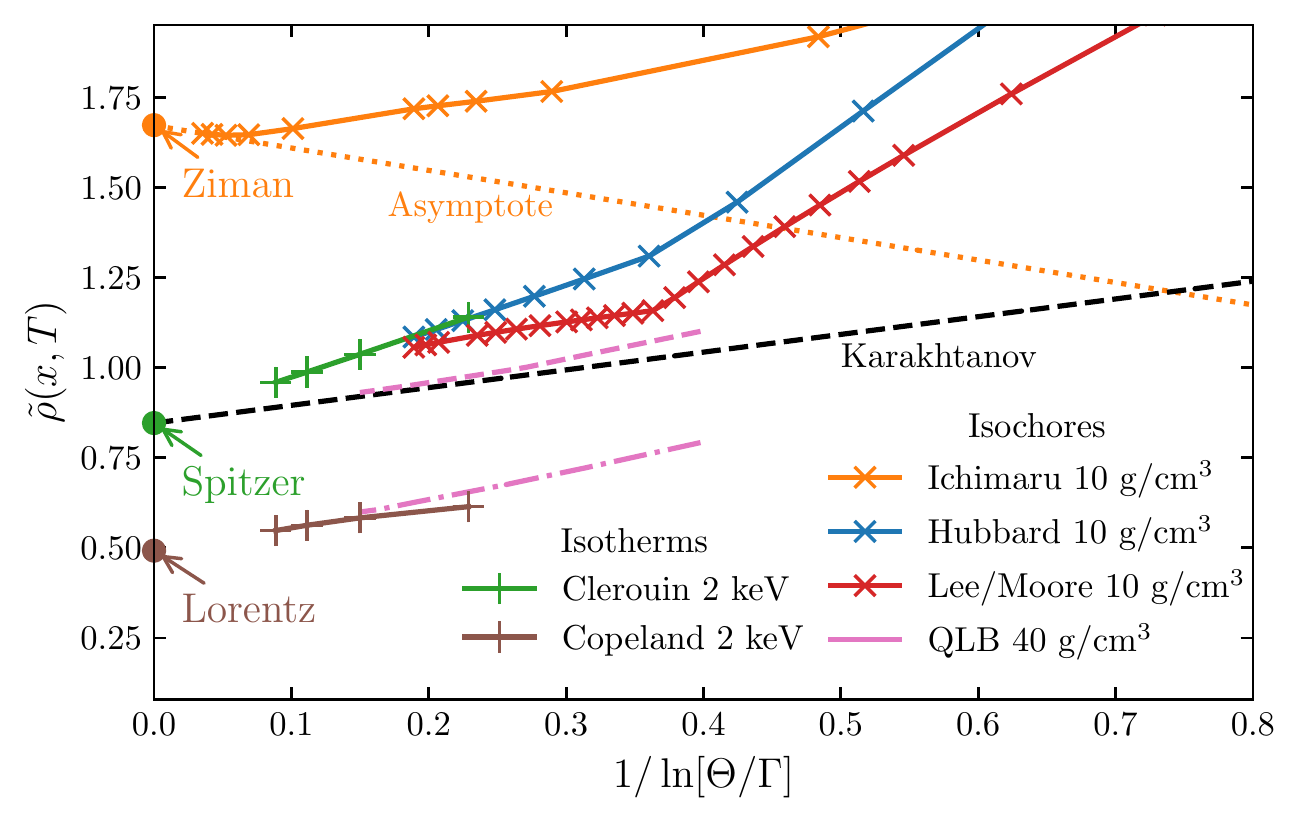}}
\caption{Analytical results for the reduced resistivity $\tilde\rho(x,T)$ (\ref{virrho}) of hydrogen plasma 
as function of $x=1/\ln (\Theta/\Gamma)$. Data for $k_BT=2000$~eV are taken from Ref.~\cite{Grabowski} 
(Cl{\'e}rouin \textit{et al.}, Copeland); lines are guide to the eye. Data for $n=10$~g/cm$^3$ are 
taken from Ref.~\cite{Lambert11} (Hubbard, Lee-More, Ichimaru). Lenard-Balescu results of 
Karakhtanov~\cite{Karachtanov16} as well as QLB results~\cite{Desjarlais} including the $e-e$ interaction 
(dashed) and without (only $e-i$: dot-dashed) are also shown. The values $\rho^{\rm Spitzer}_1$, 
$\rho^{\rm Lorentz}_1$, and $\rho^{\rm Ziman}_1$ are defined in the text. The data for the graphs are given in the 
Supplemental material~\cite{Supplementary}.}
\label{fig:1}
\end{center}
\end{figure}

\section{Virial coefficients from analytical approaches.} 
To extract the first and second virial coefficient from calculated or measured dc conductivities, we plot the expression 
\begin{equation}
 \tilde \rho(x,T) = \frac{\rho^*}{\ln (\Theta/\Gamma)}=\frac{32405.4}{\sigma (n,T) (\Omega {\rm m})} \left( T_{\rm eV} \right)^{3/2} 
 \frac{1}{\ln (\Theta/\Gamma)}
 \label{virrho}
\end{equation}
as function of $x=1/\ln(\Theta/\Gamma)$ and $T$ in Fig.~\ref{fig:1} which is denoted as \textit{virial plot}.
According to Eq.~(\ref{eq:5}), the behavior of any isotherm (fixed $T$) near $n \to 0$ is linear, 
\begin{equation}
\tilde \rho(x,T) = \tilde \rho_1(T) + \tilde \rho_2(T) x + \dots\,,
 \label{virrho1}
\end{equation}
 with $\tilde\rho_1(T)$ 
as the value at $x=0$ and $\tilde\rho_2(T)$ as the slope of the isotherm.  
\comment{As discussed above in context 
of the Born approximation (\ref{BH}), for $x > 1/\ln(100)=0.217$ the contributions of higher-order virial coefficients 
have to be taken into account. In addition, at fixed $T$, in the low-density region where $\Theta \gg 1$, the plasma is in the classical limit, and effects of degeneracy are obtained from  higher-order virial coefficients.}

In Fig.~\ref{fig:1} three cases for the first virial coefficient $\rho_1$ are shown at the axis of ordinate, see also~\cite{Roep88,RR89,Redmer97}:\\ 
i) the first virial coefficient $\rho^{\rm Spitzer}_1$ for the account of $e-e$ collisions 
according to kinetic theory (KT, Spitzer),\\ 
ii) for the neglect of $e-e$ collisions $\rho_1^{\rm Lorentz}$ 
as known from the Brooks-Herring approach for the Lorentz plasma model (BH, Lorentz),\\ 
and iii)
\begin{equation}
\rho_1^{\rm Ziman} = \frac{2}{3}(2 \pi)^{1/2}=1.67109
 \label{Ziman}
\end{equation}
for the force-force correlation function as known from the 
Ziman theory (FF, Ziman). 
In addition, the second virial coefficient $\tilde\rho_2^{\rm LB}$ of the Lenard-Balescu approximation 
(\ref{virLB}) is shown as black broken straight line which is expected to be correct in the high-temperature limit.

Two QLB calculations of Desjarlais {\it et al.}~\cite{Desjarlais} are shown in Fig.~\ref{fig:1}, see also~\cite{Supplementary}.
The  line including $e-e$ collisions obeys the same asymptote ($x \to 0$) as that of Karakhtanov~\cite{Karachtanov16}. With increasing 
$x=1/\ln(\Theta/\Gamma)$, small deviations from the linear behavior are seen. The line for calculations without $e-e$ collisions 
(Lorentz plasma) points to the corresponding asymptote given by $\rho_1^{\rm Lorentz}$.

Recently, the transport properties of hydrogen plasma were compiled in Ref.~\cite{Grabowski}. For a grid of lattice points in the $n$-$T$ plane (considering $n=0.1, 1, 10, 100$~g/cm$^3$ and $T_{\rm eV}=0.2, 2, 20, 200, 2000$) the results of different approaches were given. Large deviations were obtained which indicate not only unavoidable numerical uncertainties but also deficits in some of the theoretical approaches. Their consistency can be checked via the virial expansion as benchmark. As an example, we show data of Cl{\'e}rouin \textit{et al.} and of Copeland for the isotherm $T_{\rm eV}=2000$ taken from Ref.~\cite{Grabowski} in Fig.~\ref{fig:1}.  

Extrapolating to $x=1/\ln(\Theta/\Gamma) \to 0$, these high-temperature isotherms show already significant differences. 
The data of Cl{\'e}rouin \textit{et al.} point to the correct Spitzer limit $\rho^{\rm Spitzer}_1$, including $e-e$ collisions, but have a rather steep slope. This may be caused by the approximations in treating dynamical screening and the ionic structure factor, in contrast to a strict QLB calculation. The data of Copeland clearly point to the limit $\rho_1^{\rm Lorentz}$ of the Lorentz model, i.e., this approach does not include $e-e$ collisions and fails to describe the conductivity of hydrogen plasma correctly.

Also shown in Fig.~\ref{fig:1} are analytical results for the dc conductivity of hydrogen plasma presented in Lambert {\it et al.}~\cite{Lambert11} at the lowest density $n=10$~g/cm$^3$. The data denoted by Hubbard~\cite{Hubbard} are close to the data of Cl{\'e}rouin \textit{et al.} discussed above. The asymptote is the correct benchmark $\rho^{\rm Spitzer}_1$, but the slope is rather large.
The data of Lee and More~\cite{LeeMore} are closer to the QLB calculations. In contrast to Copeland who also claims to use the 
Lee-More approach, possibly the $e-e$ collisions are added so that the extrapolation to $x \to 0$ is near to the correct benchmark 
$\rho^{\rm Spitzer}_1$.
Because of the approximations in evaluating the Coulomb logarithm, deviations from the QLB result are seen. 
The kink in the Lee-More and Hubbard data seen in Fig.~\ref{fig:1} is due to switching the minimum impact parameter in the Coulomb logarithm from the classical distance of closest approach to the quantum thermal wave length, cf. Ref. \cite{Starrett20}.

Ichimaru and Tanaka \cite{IT85} derived an analytical expression for the conductivity which has been improved in \cite{KI95} by adding a $\tanh$-term to the Coulomb logarithm. The latter expression has been used also in Ref.~\cite{Lambert11}, the  isochore $n=10$~g/cm$^3$ is shown in Fig.~\ref{fig:1}.
The approach is based on a single Sonine polynomial approximation where the effect of $e-e$ collisions is not taken into account. 
The empirical fit of Kitamura and Ichimaru~\cite{KI95} approximates the conductivity for degenerate plasmas, see also Fig.~9 of Ref.~\cite{Lambert11}. 
However, in the low-density limit this approach fails to describe the conductivity approaching $\rho^{\rm Ziman}_1$ at $x=0$.

\section{Virial representation of DFT-MD simulations.} 
\comment{DFT-MD simulations are of great interest, since they do not}
suffer from the restrictions of perturbation theory as typical for analytical results 
and can directly be confronted with the virial expansion. In addition, with the virial expansion the results 
can be extrapolated to the low-density region were DFT-MD simulations become infeasible. 

In this work, we present new DFT-MD results for the electrical conductivity \comment{of hydrogen} obtained from an evaluation of the Kubo-Greenwood formula~\cite{Kubo66,Greenwood,French2017,Gajdos2006}. The \comment{125-atom} simulations are performed with the Vienna \textit{ab initio} simulation package (VASP)~\cite{Kresse1993,Kresse1994,Kresse1996} using the exchange-correlation functional of Perdew, Burke, and 
Ernzerhof (PBE)~\cite{Perdew1996} and the provided Coulomb potential for hydrogen. 
The time steps were chosen between 0.2 and 0.1~fs and the simulations ran for at least 4000 time steps. The ion temperature is controlled with a \comment{Nos{\'e}-Hoover} thermostat \cite{Nose}.
For all simulations, the reciprocal space was sampled at the Baldereschi mean value point~\cite{Baldereschi1973}.
Special attention has been paid to convergence \comment{with respect to} the particle number.
\comment{Additional details of the simulations} are given in the supplemental material \comment{and the results are given in Tab.\ref{tab:Max1}.} 


\begin{table}[htp]
\caption{\comment{Virial representation of the dc conductivity $\sigma$ and of 
 $\tilde \rho(x,T)$ with $x=1/\ln(\Theta/\Gamma)$: the values for $\sigma$ 
 and $\tilde \rho$ result from our own DFT-MD simulations (this work; number 
 of atoms: 125). Only positive values for $x=1/\ln(\Theta/\Gamma)$, i.e.\ 
 $\Theta/\Gamma > 1$, are considered.}}
\begin{center}
\begin{tabular}{|c|c|c|c|c|c|c|}
\hline
 $n$ & $k_BT$ & $\Gamma$ & $\Theta$& $1/\ln(\Theta/\Gamma)$ & $\sigma$ &$\tilde \rho(x,T)$\\
{[}g/cm$^3$] &[eV] &  & &  & [MS/m]  &\\
\hline
 2  & 50  	& 0.49275 	& 1.2172 	&1.1059		  &  7.170  & 1.767   \\
 2  & 75 	  & 0.3285 		& 1.8257 	&0.58302		&  11.44  & 1.073   \\
 2  & 100  	& 0.24637 	& 2.4343 	&0.43657 		&  15.26  & 0.9269 \\ 
 3  & 100 	& 0.28203 	& 1.8577 	&0.53047 		&  16.85  & 1.020  \\ 
 3  & 150  	&0.18802 		&2.7866 	&0.37092 		&  25.67  & 0.8603 \\ 
 4  & 150 	&0.20694 		&2.3003 	&0.41522 		&  27.39  & 0.9026 \\ 
\hline
\end{tabular}
\end{center}
\label{tab:Max1}
\end{table}

\begin{figure}[tbp]
\begin{center}
\centerline{\includegraphics[width=\columnwidth,angle=0]{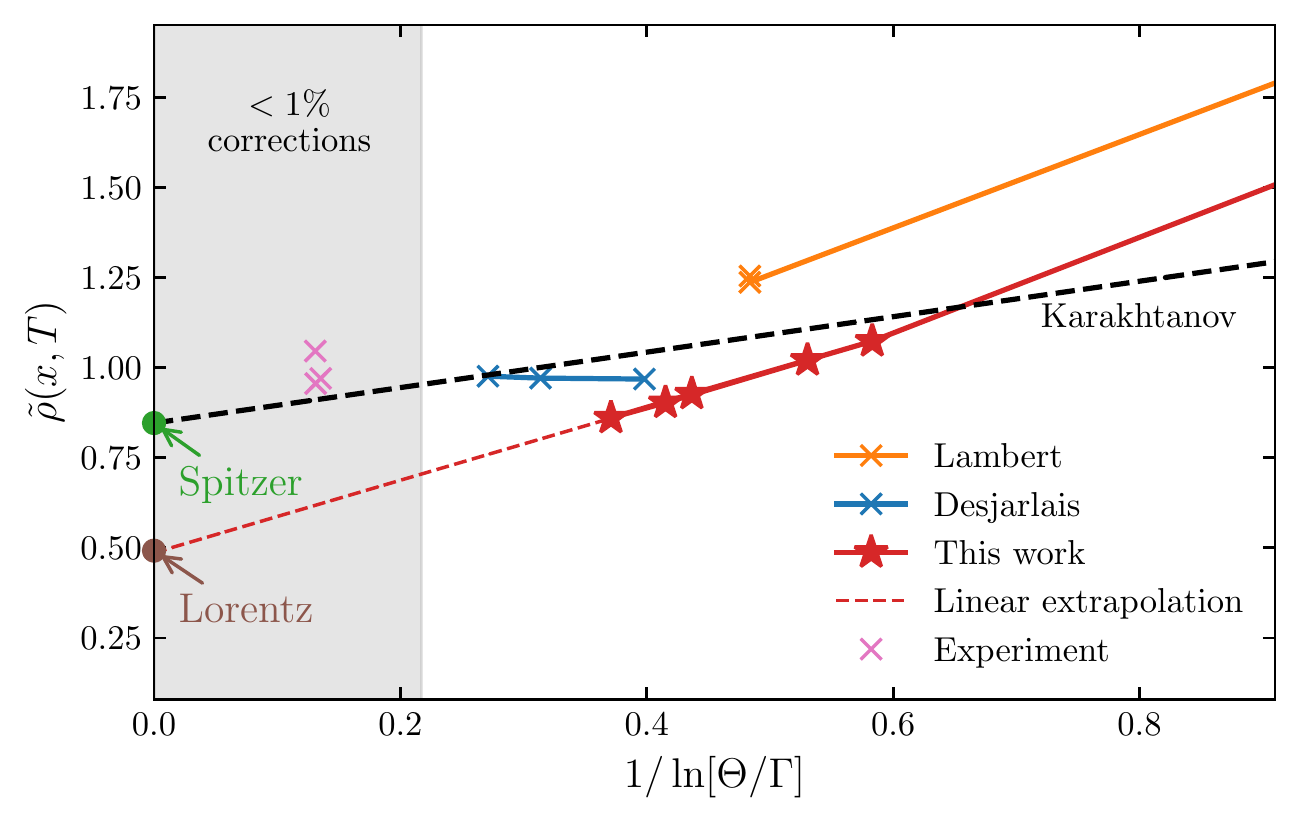}}
\caption{Reduced resistivity $\tilde\rho(x,T)$ (\ref{virrho}) for hydrogen plasma as a function of $x=1/\ln (\Theta/\Gamma)$: 
QMD simulations of Lambert {\it et al.} \cite{Lambert11} (the orange line points to the value for $n=80$ g/cm$^3$, $k_BT=300$ eV), DFT-MD simulations of Desjarlais \textit{et al.} \cite{Desjarlais} and of this work, experimental values of G{\"u}nther and Radtke~\cite{Guenther}, and Lenard-Balescu results of Karakhtanov~\cite{Karachtanov16}.  
$\rho^{\rm Spitzer}_1=0.846$ and $\rho^{\rm Lorentz}_1=0.492$ are defined in the text.
The shaded area indicates the region where corrections to the linear behavior of the virial expansion amount to less than 1\%. The red dashed line represents a linear extrapolation of our results based on a linear fit to our three leftmost results.
Data are given in the Supplemental material \cite{Supplementary}. 
}
\label{fig:DFTcond}
\end{center}
\end{figure}

\comment{Our DFT-MD results are plotted in Fig.~\ref{fig:DFTcond} and show a general increase with an increasing $x=1/\ln (\Theta/\Gamma)$. In comparison, the virial plot contains previous DFT-MD conductivity data \cite{Lambert11, Desjarlais}, which were 
translated into our $\tilde \rho$ framework. The first set of previous DFT-MD calculations} has been published by Lambert {\it et al.}~\cite{Lambert11} which were also used by Starrett~\cite{Starrett16}. 
Results for $\tilde \rho$ for the lowest values of $x>0$ at three different densities are given in Fig.~\ref{fig:DFTcond}.
Inspecting Fig.~\ref{fig:DFTcond}, values for 10~g/cm$^3$ at 200~eV and for 160~g/cm$^3$ at 800~eV are 
close together, i.e., we see a dominant dependence on $x$, no additional density or temperature effect is seen. 
They are also close to the Lee-More approach including $e-e$ collisions so that they are not in conflict with
the correct benchmark (KT, Spitzer). Calculations are based on a formulation of the Kubo-Greenwood method 
for average atom models neglecting the ion structure factor~\cite{Starrett12} so that these QMD values are 
possibly also influenced by approximations and, therefore, deviate slightly from other calculations. However, 
the parameter values $x$ are too large to estimate the virial expansion.

\comment{The second set of previous} DFT-MD simulations for hydrogen plasma in the low-$x$ region were given by Desjarlais {\it et al.}~\cite{Desjarlais}, 
see Fig.~\ref{fig:DFTcond}. For a density of 40~g/cm$^3$, three temperatures $T_{\rm eV} = 500, 700$, and $900$ were 
considered. The reduced resistivity $\tilde \rho_1(x,T)$ approaches the benchmark obtained from the QLB calculations. 
However, the linear extrapolation to $\rho^{\rm Spitzer}_1$ at $x=0$ is not seen by these data.

Interestingly, the results for $\tilde \rho$ of the different DFT-MD simulations \comment{do not} follow approximately a single curve as expected from the high-temperature limit of the virial expansion. The values of Lambert {\it et al.} are significantly \comment{higher} than ours but the slope is almost the same. While we employ the generalized gradient approximated exchange-correlation energy of PBE~\cite{Perdew1996}, Lambert \textit{et al.} use the local density approximation. 
\comment{They used orbital-free molecular dynamics in order to simulate the system and obtain various snapshots for each density-pressure point. Subsequently, these snapshots were evaluated via the Kubo-Greenwood formula using the Kohn-Sham code Abinit, which is equivalent to our used VASP implementation.}
The DFT-MD simulations of Desjarlais {\it et al.}~\cite{Desjarlais} are close to our results, but the slope of the virial plot is quite different. 
DFT-MD simulations are usually performed at high densities where the electrons are degenerate so that $e-e$ collisions can be neglected. In the low-density region ($x < 1$) as considered here, we could improve the accuracy by studying the convergence of the DFT-MD results, in particular with respect to particle number and cut-off energy, using high-performance computing facilities.

A long-discussed problem in this context is the question whether or not $e-e$ collisions are taken into account within the DFT-MD formalism. For example, in Ref.~\cite{Heidi} it was pointed out that a mean-field approach is not able to describe two-particle correlations, in particular $e-e$ collisions. However, $e-e$ interaction is taken into account by the exchange-correlation energy as shown in Ref.~\cite{Desjarlais} by comparing DFT-MD data for the electrical conductivity to QLB results. 
The calculations of Desjarlais {\it et al.}~\cite{Desjarlais} for $n=40$~g/cm$^3$ and our present ones for $n=2$~g/cm$^3$ were computationally demanding but are still not very close to $x=0$ so that extrapolation to the limit $x=0$ is not very precise. 
However, the corresponding slopes are quite different: while the present DFT-MD data favor 
$\rho^{\rm Lorentz}_1$ as asymptote at $x=0$, those of Ref.~\cite{Desjarlais} seem to point
to the Spitzer value, Eq.~(\ref{eq:Spitzer}).
Thus, our results do not solve the lively debate on whether or not DFT-MD simulations
include the effect of $e-e$ collisions on the conductivity or not. 
We conclude that further DFT-MD simulations have to be performed for still higher 
temperatures and/or lower densities in order to approach the limit $x \to 0$ so that
the value for $\rho_1$ can be derived more accurately. Such simulations, \comment{e.g. for densities below 1 g/cm$^3$,} are computationally 
very challenging using the Kohn-Sham DFT-MD method so that alternative schemas like 
stochastic DFT~\cite{Cytter} or the spectral quadrature method~\cite{Sharma} have to 
be applied for this purpose.

\comment{We would like to mention that in the case of thermal conductivity it has been shown that 
the contribution of $e-e$ collisions is not taken into account in DFT-MD simulations \cite{Desjarlais}
and gives an additional term. 
A profound  discussion  on the mechanism of $e-e$ collisions
has been given recently by Shaffer and Starrett \cite{Starrett20}. 
They argued that the precise nature of the incomplete account of $e-e$ scattering may be resolved by methods going
beyond the Kubo-Greenwood approximation such as time-dependent DFT or GW corrections. 
Considering a quantum Landau-Fokker-Planck kinetic theory,
their main issue is that scattering between particles in a plasma should be described not by the Coulomb interaction but by the potential 
of mean force. Obviously, if part of the interaction is already taken into account introducing quasiparticles
and mean-field effects, the corresponding contributions must be removed from the  Coulomb interaction for $e-e$ scattering 
to avoid double counting.
Comparing with QMD results, Shaffer and Starrett \cite{Starrett20} point out that their findings support the conclusions of Ref. \cite{Desjarlais}
that the Kubo-Greenwood QMD calculations contain the indirect electron-electron reshaping effect relevant to both the electrical
and thermal conductivity, but they do not contain the direct scattering effect which further reduces the thermal conductivity.}

\section{Experiments.} 
Ultimately, the virial expansion (\ref{eq:vir}) has to be checked experimentally but 
accurate data for the conductivity of hydrogen plasma in the low-density limit and/or 
at high temperatures are scarce. Accurate conductivity data for dense hydrogen plasma 
were derived by G{\"u}nther and Radtke~\cite{Guenther} which are shown in the virial 
plot, Fig. \ref{fig:DFTcond}. They are close to the benchmark data of the virial expansion.
Note that systematic errors are connected with the analysis of such experiments. For 
instance, the occurrence of bound states requires a realistic treatment of the plasma 
composition and of the influence of neutrals on the mobility of electrons. Alternatively, 
conductivity measurements in highly compressed rare gas plasmas have been performed by 
Ivanov {\it et al.}~\cite{Ivanov} and Popovic  {\it et al.}~\cite{Popovic,Esser03}, 
but the interaction of the electrons with the ions deviates from the pure Coulomb 
potential owing to the cloud of bound electrons. The corresponding virial plot is close 
to the data of hydrogen plasma, see \cite{Supplementary}, but requires a more detailed 
discussion with respect to the role of bound electrons.

\section{Conclusions.} 
We propose an exact virial expansion (\ref{eq:vir}) for the plasma conductivity to analyze 
the consistency of theoretical approaches. For instance, several analytical calculations of 
the dc conductivity $\sigma(T,n)$ presented in Ref.~\cite{Grabowski} miss this strict 
requirement and fail to give accurate results. Results of DFT-MD simulations are presently 
considered to be most reliable, and future PIMC simulations can be tested by benchmarking 
with the virial expansion (\ref{eq:vir}) for $x \to 0$. Note that these \textit{ab initio} 
simulations become computationally challenging in the low-density region, but the virial 
expansion allows the extrapolation into this region. The construction of interpolation 
formulas is possible, see~\cite{Esser03}, if the limiting behavior for $n \to 0$ and further 
data in the region of larger densities not accessible for analytical calculations are known. 

\comment{An outstanding problem that could potentially be addressed by applying the virial expansion of the conductivity is} the question 
whether or not the $e-e$ collisions are rigorously taken into account. 
\comment{Despite the work presented in \cite{Desjarlais,Starrett20}, }\comment{there is no final proof whether the Kubo-Greenwood QMD calculations with the standard expressions for the exchange-correlation
energy functional give the exact value for the plasma conductivity in the low-density limit. A Green's function approach may
solve this problem but this has not been performed yet.} 
\comment{Therefore, we suggest to apply our benchmark criterium on future large data sets of Kubo-Greenwood QMD calculations to investigate the contribution of
 $e-e$ collisions in the low-density limit.}

The approach described here is applicable also to other transport properties such as thermal conductivity, thermopower, viscosity, and diffusion coefficients. 
Of interest is also the extension of the virial expansion to elements other than hydrogen, where different ions may be formed and the electron-ion interaction is no longer pure Coulombic. \\

\newpage
{\bf Acknowledgments}\\

We thank M.~Desjarlais, M.~French, and V.~Recoules for valuable and fruitful discussions \comment{and for providing data sets}. 
This work was supported by the North German Supercomputing Alliance (HLRN) and the ITMZ of the University of Rostock.
MS and RR thank the DFG for support within the Research Unit FOR~2440. 
MB was supported by the European Horizon 2020 program within the Marie Sk{\l}odowska-Curie actions (xICE, grant number 894725).



\begin{thebibliography}{10}
\bibitem{foot1}
Improving the static screening which gives for $\tilde \rho_2$ the Debye value $-0.814047$, see Eq.~(\ref{BH}), dynamical screening has been considered by \cite{GDW,WDW} with the result $-0.4129$. Based on generalized linear response theory (Zubarev approach), the treatment of dynamical screening solving the quantum Lenard-Balescu equation \cite{RR89,RRMK89,RRT89} gives $0.40668$, also used in \cite{Esser03}. The value (\ref{virLB}) was calculated with higher precision in \cite{Karachtanov13}.
\end{thebibliography}
\end{document}